\documentclass[11pt,twoside]{article}
\usepackage{asp2010}

\resetcounters

\newcommand{\logg}{\mbox{$\log g$}}
\newcommand{\loggw}[1]{\mbox{$\log g\hspace{-0.5mm} =\hspace{-0.5mm}  #1$}}
\newcommand{\Teff}{\mbox{$T_\mathrm{eff}$}}
\newcommand{\Teffw}[1]{\mbox{$\Teff\hspace{-0.5mm}=\hspace{-0.5mm}#1\,\mathrm{K}$}}
\newcommand{\sga}{\mbox{\raisebox{-0.10em}{$\stackrel{>}{{\mbox{\tiny $\sim$}}}$}}}

\newcommand{\sla}{\raisebox{-0.10em}{$\stackrel{<}{{\mbox{\tiny $\sim$}}}$}}

\newcommand{\aador}{AA\,Dor}

\begin{document}

\title{Spectral Analysis within the Virtual Observatory:\\ The GAVO Service \emph{TheoSSA}}
\author{Ellen Ringat
\affil{Institute for Astronomy and Astrophysics,
       Kepler Center for Astro and Particle Physics,
       Eberhard Karls University,
       Sand 1,
       D-72076 T\"ubingen,
       Germany\\email: gavo@listserv.uni-tuebingen.de}
}

\begin{abstract}
In the last decade, numerous Virtual Observatory organizations were established. 
One of these is the German Astrophysical Virtual Observatory (\emph{GAVO}) that 
e.g\@. provides access to spectral energy distributions via the service \emph{TheoSSA}. 
In a pilot phase, these are based on the T\"ubingen NLTE Model-Atmosphere Package (\emph{TMAP}) 
and suitable for hot, compact stars. We demonstrate the power of \emph{TheoSSA} in an application 
to the sdOB primary of AA Doradus by comparison with a ``classical" spectral analysis. 
\end{abstract}

\section{Introduction}
\label{sect:introdcution}

With the progress in observational and computer technology, 
the quantity of astronomical data was and is increasing exponentially. 
To be able to handle the actual and future amount of data, Virtual Observatories (VOs) were established. 
Their aims are to make the data globally accessible, to prevent losing data, and to develop tools, 
services, and standards to enable research using the internet. 
The idea of the VO is rather old, starting with the first databases and tools 
such as e.g\@. 
SIMBAD\footnote{http://simbad.u-strasbg.fr/simbad/, http://aladin.u-strasbg.fr/aladin.gml, http://iraf.noao.edu/} (1979), 
ALADIN$^1$ (1999), or 
IRAF$^1$ (1984). 
Since 2000 more and more national VOs were founded to make astronomical data accessible. 
Many of them joined the International Virtual Observatory Alliance\footnote{http://www.ivoa.net/} (IVOA). 
Observational and theoretical data sets were prevented from getting lost by making them public. 
Standards were defined to get a uniform shape of all data. Services and tools for data-handling were developed, 
and contact points were created to aid publishing data and using tools and services. 

One of these national organizations is the German Astrophysical Virtual Observatory (\emph{GAVO}\footnote{http://g-vo.org/}), 
the German effort to contribute to this idea. Within the framework of a \emph{GAVO} project, 
the well-established 
T\"ubingen NLTE Model-Atmosphere Package 
\citep[\emph{TMAP},\footnote{http://astro.uni-tuebingen.de/\raisebox{.3em}{\tiny $\sim $}TMAP}][]{werneretal_2003} 
was made accessible. This model-atmosphere code and the different ways to access
its products are described in the following sections.

\section{The Basis - The T\"ubingen NLTE Model-Atmosphere Package}
\label{sect:tmap}

\emph{TMAP} is used to calculate stellar model atmospheres in hydrostatic and radiative equilibrium. 
It assumes plane-parallel geometry due to the relatively thin atmosphere of a hot, compact star compared 
to its radius. Up to about 1500 atomic levels can be treated in NLTE, about 4000 lines for classical model 
atoms and more than 200 millions of lines for iron-group elements (Ca-Ni) can be considered. Currently all 
elements from hydrogen to nickel can be considered \citep{rauch_2003}. For the iron-group elements the program 
\emph{IrOnIc} \citep{rauchdeetjen_2003} is used to calculate sampled cross-sections and to set up atomic data files. 
This program is necessary to handle the hundreds of millions of lines of iron-group elements and to reduce them 
with a statistical treatment to a number suitable for the model atmosphere code.

\emph{TMAP} is used for hot, compact objects with 
$T_\mathrm{eff} \ \sga\  20000\,$K and $4 \ \sla\  g \ \sla\  9$. 
The \emph{GAVO} service \emph{TheoSSA} is based on this model-atmosphere code at the moment.

\section{Implementation - The VO Service \emph{TheoSSA}}
\label{sect:theossa}

\begin{figure}[ht]
\setlength{\textwidth}{13.85cm}
\plotone{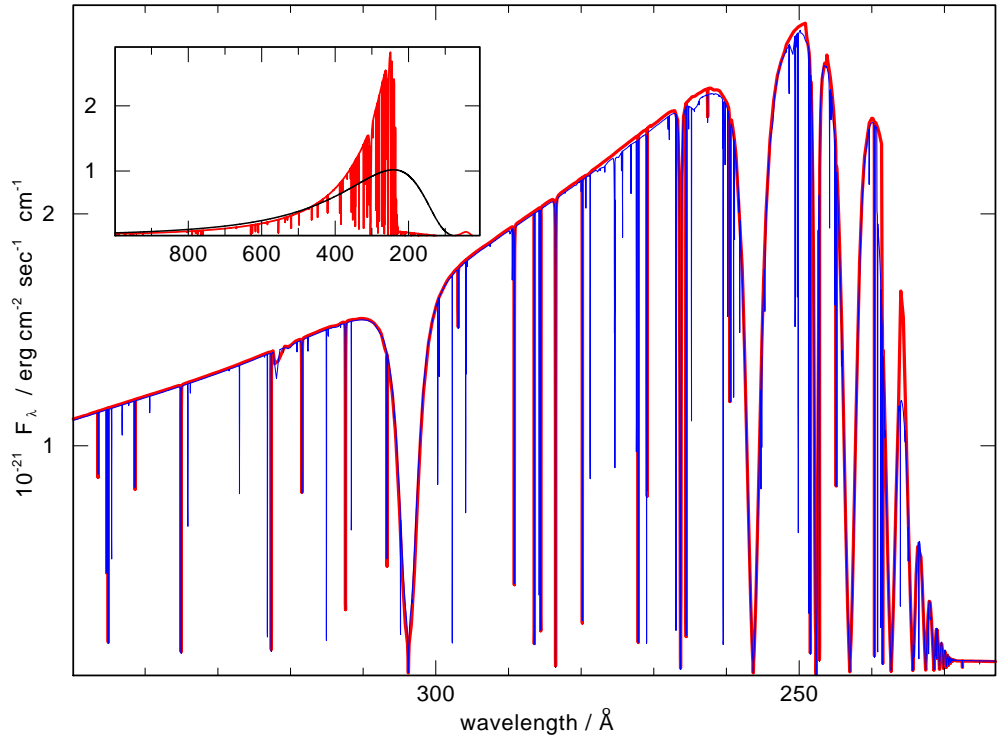}
\caption{Comparison of the 
         \emph{TMAP} (thin line, blue in the online version) and 
         \emph{TMAW} (thick, red) fluxes (\Teffw{120000}, \loggw{7.5}). 
         The strongest deviations can be seen in this region, but still they are very small compared to a
         blackbody flux (black) with the same $T$ (insert).}
\label{fig:bb}
\end{figure}

\emph{TheoSSA}\footnote{http://dc.g-vo.de/theossa} was originally planned to 
provide spectral energy distributions (SEDs) as input for photoionization models for planetary nebulae (PNe). 
For these models, blackbodies were used very often and, thus, such PN analyses yielded wrong results. 
The blackbody approximation differs vastly from our model fluxes (Fig.\,\ref{fig:bb}). In order to supply 
reliable ready-to-use SEDs this service gives access in three ways.\vspace{2mm}\\
\hbox{}\hspace{5mm}1) download already calculated SEDs \\
\hbox{}\hspace{5mm}2) calculate individual SEDs with the simulation software 
\emph{TMAW}\footnote{http://astro.uni-tuebingen.de/\raisebox{.3em}{\tiny $\sim$}TMAW} \\
\hbox{}\hspace{5mm}3) download ready-to-use atomic data or tailor individual model atoms with 
\emph{TMAD}\footnote{http://astro.uni-tuebingen.de/\raisebox{.3em}{\tiny $\sim $}TMAD} \\

The \emph{TheoSSA} database is controlled via a web interface where the fundamental parameters 
($T_\mathrm{eff}, \log g$, abundances) are entered. Then all available SEDs within a chosen parameter range 
are displayed and can be downloaded. The database is growing in time, because SEDs calculated with 
\emph{TMAW} are automatically ingested. In case that the calculation procedure is significantly changed, 
these SEDs are re-calculated with the latest \emph{TMAW} version.

\section{More Detailed/Individual SEDs - \emph{TMAW}}
\label{sect:tmaw}

\emph{TMAW} is the simulation software of the model-atmosphere code. Like \emph{TheoSSA}, it is 
controlled via web interface where the requested parameters and the email address are entered.
The sending process starts the calculation of a SED with the requested parameters on the PC cluster
of the Institute for Astronomy and Astrophysics T\"ubingen or, if the number of requests is $ > 150$, 
on compute resources of \emph{AstroGrid-D}\footnote{http://www.gac-grid.net/}.

\begin{figure}[ht]
\setlength{\textwidth}{8cm}
\plotone{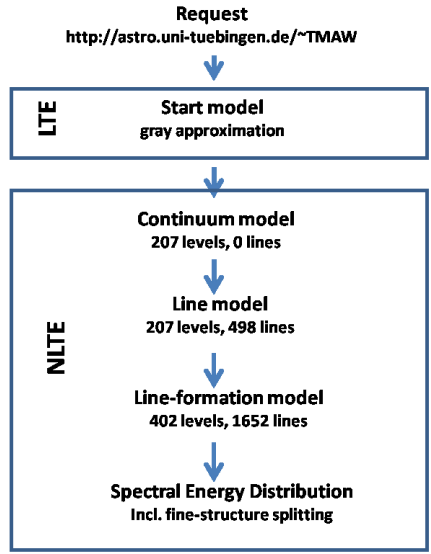}\vspace{-3mm}
\caption{Standard procedure of a \emph{TMAW} calculation. 
The numbers are taken from the example given in the text.}
\label{fig:Schema}
\end{figure}

In the standard procedure, the steadily updated, ready-to-use model atoms from \emph{TMAD} are 
taken as input and, after the LTE start model is created, a continuum model is calculated (Fig.\,\ref{fig:Schema}). 
Then lines are inserted initially without temperature correction, then with Uns\"old-Lucy temperature correction 
\citep{unsoeld_1968}. After about one or two days and at least 300 iterations, line profiles are calculated and the 
result is sent to the email address of the user.

At the moment, this service can consider the elements hydrogen, helium, carbon, nitrogen, and oxygen, 
but in the future the next most abundant elements, neon and magnesium (as well as the iron-group elements, Ca - Ni), will be included.

\section{Creating The Model Atoms - \emph{TMAD}}
\label{sect:tmad}

\begin{figure}[ht]
\setlength{\textwidth}{13.85cm}
\plotone{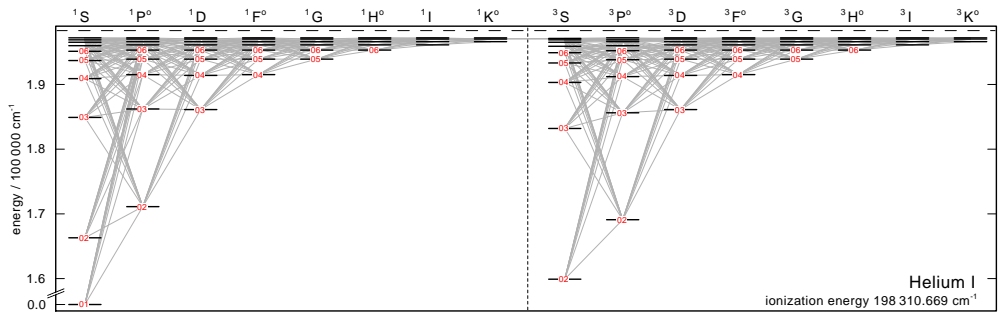}
\caption{Grotrian diagram of HE\,\textsc{i} (as provided by \emph{TMAD}), 
         singlet (left) and triplet (right) systems.}
\label{fig:HE}
\end{figure}

The T\"ubingen Model Atom Database (\emph{TMAD}) was originally created for \emph{TMAP} and 
therefore its format is compatible with it. It is updated regularly. For example the He\,{\sc i} 
model atom was enlarged in the latest update due to recently published atomic data retrieved from NIST. 
Now levels up to a principal quantum number of $n=10$ instead of $n=5$ can be considered as NLTE levels (Fig.\,\ref{fig:HE}).

\begin{figure}[ht]
\setlength{\textwidth}{7.0cm}
\plotone{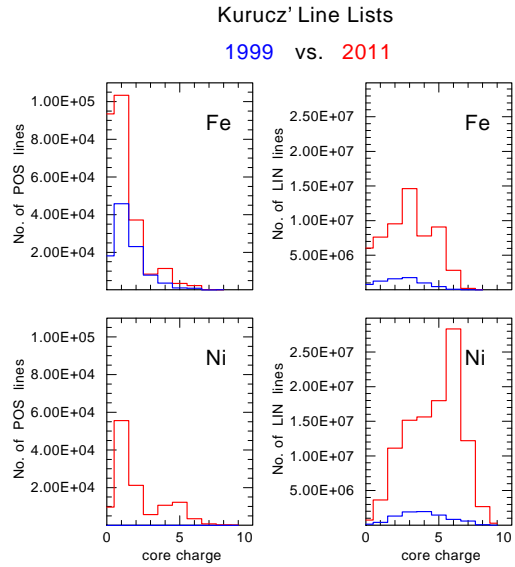}
\caption{Comparison of the number of POS (left) and LIN (right) lines for Fe (top) and Ni (bottom) 
         before and after the extension of Kurucz' line lists.}
\label{fig:LINPOS}
\end{figure}

Presently, the elements H, He, C, N, O, F, Ne, Na, Mg, Si, S, Ar, and Ca are included. 
Main data sources for \emph{TMAD} are the standard atomic databases like 
NIST\footnote{http://www.nist.gov/pml/data/asd.cfm} 
or the Opacity Project\footnote{http://cdsweb.u-strasbg.fr/topbase/op.html}.
 
\emph{TMAD} supplies two kinds of data. Level energies and radiative as well as collisional 
transitions can be found for model atmosphere and for SED calculations (the latter including fine-structure splitting). 
For model atmosphere calculations ready-to-use or individual data can be downloaded. This data is 
accessible to the community and can be used for all desired purposes, e.g\@. as input for another model-atmosphere code.

\section{Handling Iron-Group Elements - \emph{IrOnIc}}
\label{sect:ironic}

Another \emph{GAVO} project is the parallelization of the \emph{IrOnIc} code \citep{rauchdeetjen_2003}. 
Presently, it may take one to three days to calculate cross-sections and a model atom with \emph{IrOnIc}. 
To accelerate, we write a new version of this code using MPI as well as GPU (CUDA) parallelization. 
This service will be controlled via a web interface in a similar way to \emph{TMAW} and will be accessible to the public.

Input for the \emph{IrOnIc} code are 
Kurucz' line lists\footnote{http://kurucz.harvard.edu/atoms.html}
These contain LIN and POS lines. The latter are only lines with measured (good) wavelengths, LIN lines
have theoretically calculated entries additionally and therefore about 300 times more lines than POS lists. 
New data became available recently 
\citep[Fig.\,\ref{fig:LINPOS},][]{kurucz_2009}. 
Additional data sources are the
OPACITY and IRON projects \citep{seatonetal_1994,hummeretal_1993}

\begin{figure}[ht]
\setlength{\textwidth}{13.85cm}
\plotone{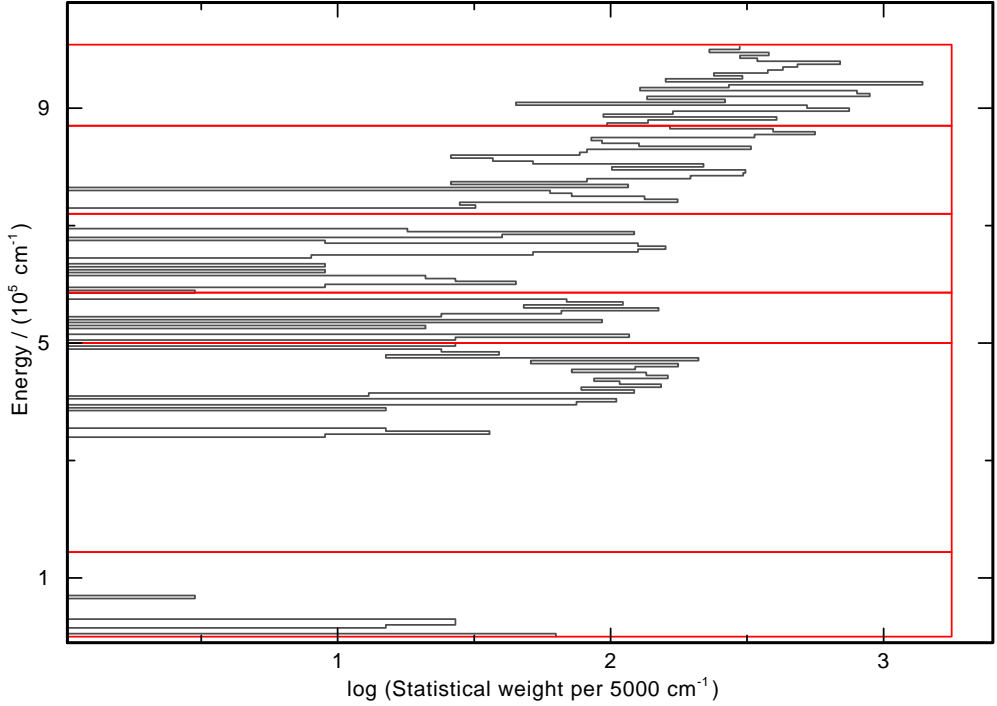}
\caption{Energy range of Fe\,\textsc{vii} divided into bands to create superlevels. 
         The ionization energy is $1.008 \cdot 10^{6}\,\mathrm{cm^{-2}}$.}
\label{fig:bands}
\end{figure}

\begin{figure}[ht]
\setlength{\textwidth}{13.85cm}
\plotone{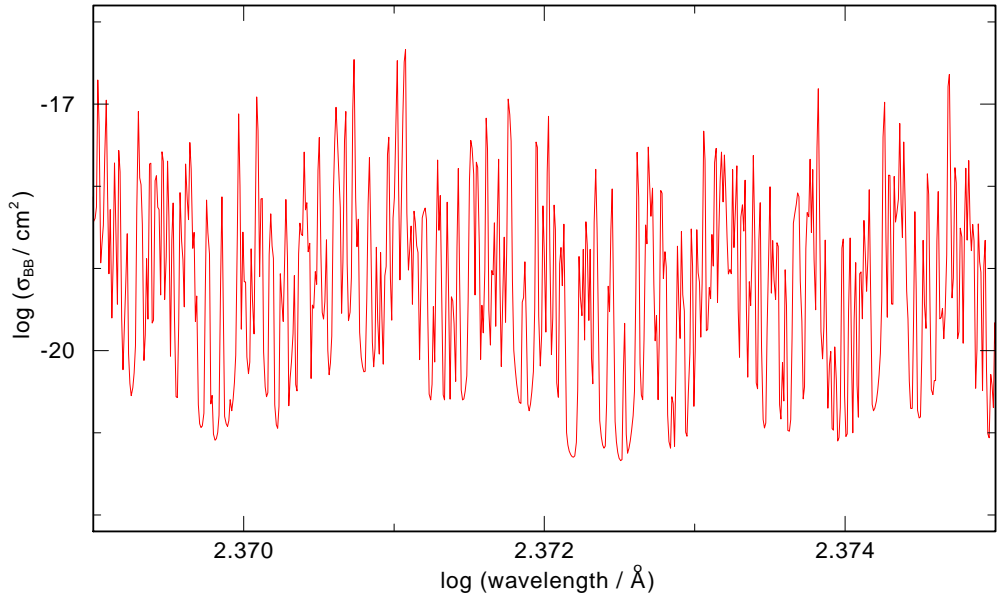}
\caption{Sampled RBB cross-sections for Fe\,\textsc{vii} (example transition between our bands 2 and 5,
         Fig.~\ref{fig:bands}).}
\label{fig:RBB}
\end{figure}

\emph{IrOnIc} divides the energy range up to the ionization energy of an ion into several bands 
(Fig.\,\ref{fig:bands}). All levels within one band contribute to this so-called superlevel via defining 
its energy and statistical weight. This superlevels are treated in NLTE, the levels within one band are 
in LTE relation. Therefore the number of NLTE levels is reduced to a number manageable for the model 
atmosphere code. For all transitions within one band or between different bands, the cross-sections 
are sampled (Fig.\,\ref{fig:RBB}).

\hbox{~}\vspace{5mm}
\section{A First Benchmark Test - The Case of AA\,Doradus}
\label{sect:aador}

The initial aim of \emph{TheoSSA} and \emph{TMAW} was to 
provide easy access to synthetic stellar fluxes (Sect.\,\ref{sect:theossa})
for the PNe community, which was interested in realistic stellar ionizing fluxes
for their PN models. It turned out rather quickly, that other groups were interested in
X-ray, optical, and infrared fluxes as well. \emph{TheoSSA} / \emph{TMAW} were then
extended for this purpose. While the accuracy in ionizing fluxes is better than
10\,\% compared to \emph{TMAP} (Sect.\,\ref{sect:tmap}) calculations with the most detailed model atoms, 
the precision in the optical is limited due to the smaller, standard model atoms
that are used in the \emph{TMAW} model-atmosphere calculations. The aim is here
to perform preliminary spectral analyses with \emph{TMAW} SEDs and
to be better than 20\,\% in \Teff, \logg, and abundance determinations.

Publishing data of all kind is a very important task of the VO.
For all these data, quality control should be a pre-requisite and is highly desirable.
Therefore it is necessary to establish benchmark tests for VO data, services, and tools. 
This is essential for theoretical data and simulations as well.

\subsection{\aador}
\label{sect:aador}

The eclipsing, post common-envelope binary \aador\ has an sdOB star primary star whose optical 
spectrum shows the broad Balmer lines and the He\,{\sc ii}\,$\lambda$ 4686\AA\, 
line, typical for this class. A recent analysis of the primary \citep{klepprauch_2011}
yielded \Teffw{42000} and \loggw{5.46} within small error limits. 
For this analysis, an extended grid of NLTE model atmospheres 
considering opacities of the elements 
H, He, C, N, O, Mg, Si, P, S, Ca, Sc, Ti, V, Cr, Mn, Fe, Co, and Ni was calculated.
\Teff\ and \logg\ were then determined by a $\chi^2$-fit method
based on synthetic spectra of these models and
high-resolution and high-S/N optical UVES observations
(cf\@. Rauch these proceedings).

\begin{figure}[ht]
\setlength{\textwidth}{13cm}
\plotone{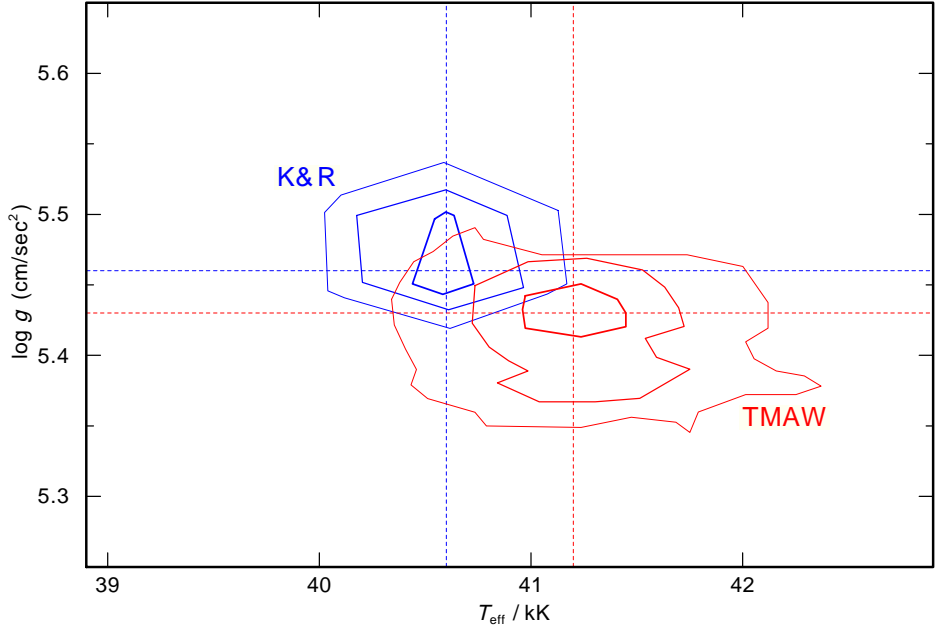}
\caption{Formal 1\,$\sigma$, 2\,$\sigma$, and 3\,$\sigma$
         contour lines of the $\chi^2$ fits in the \Teff\ - \logg\ plane.
         K\&R denotes the detailed models of \citet{klepprauch_2011},
         TMAW the \emph{TMAW} models.
         }
\label{fig:chi}
\end{figure}

\aador\ is, thus, an ideal object to prove the \emph{TMAW} service for its adequacy for spectral analysis.
For this test, we recalculated the fine grid of \citet{klepprauch_2011} via
\emph{TMAW} (Fig.\,\ref{fig:Schema}). 
We used their abundances, 
\Teffw{39500 - 43500} with $\Delta T_\mathrm{eff}  = 250\,$K, and
\loggw{5.30 - 5.60}, with $\Delta \loggw{0.01}$.
In total, 527 HHeCNO-composed models were calculated in one week on compute resources of 
\emph{AstroGrid-D}. Analogously to \citet{klepprauch_2011}, we calculated synthetic
spectra from these models and performed a $\chi^2$ fit to the 
UVES\footnote{Ultraviolet and Visual Echelle Spectrograph at ESO's VLT}
observations (ProgId 66.D$-$180). 
Astonishingly, the result is very good (Fig\@.~\ref{fig:chi}).
We arrive at \Teffw{41150} and \loggw{5.43} while the
$\chi^2$ fit of \citet{klepprauch_2011} results in
\Teffw{40600} and \loggw{5.46}. This are deviations of about 1\,\% in \Teff\ and
7\,\% in \logg, which is better than our expectation. The wider and irregular
contour lines of the \emph{TMAW}-based $\chi^2$ fit (Fig\@.~\ref{fig:chi}) are
due to the fact that only about 300 iterations were performed in contrast to
about 8000/model in the grid of \citet{klepprauch_2011} and, thus, the \emph{TMAW} models are
not homogeneously converged.

\newpage

\subsection{Conclusion and further improvements}
\label{sect:improvements}

The example of \aador\ shows clearly, that \emph{TMAW} models are well suited for the
spectral analysis of optical spectra. The calculation times for grids via
\emph{TMAW} are small because of the \emph{AstroGrid-D} computational power in the background.
However, with some improvements, we enhanced the reliability of \emph{TMAW} models.

\begin{figure}[ht]
\setlength{\textwidth}{13.85cm}
\plotone{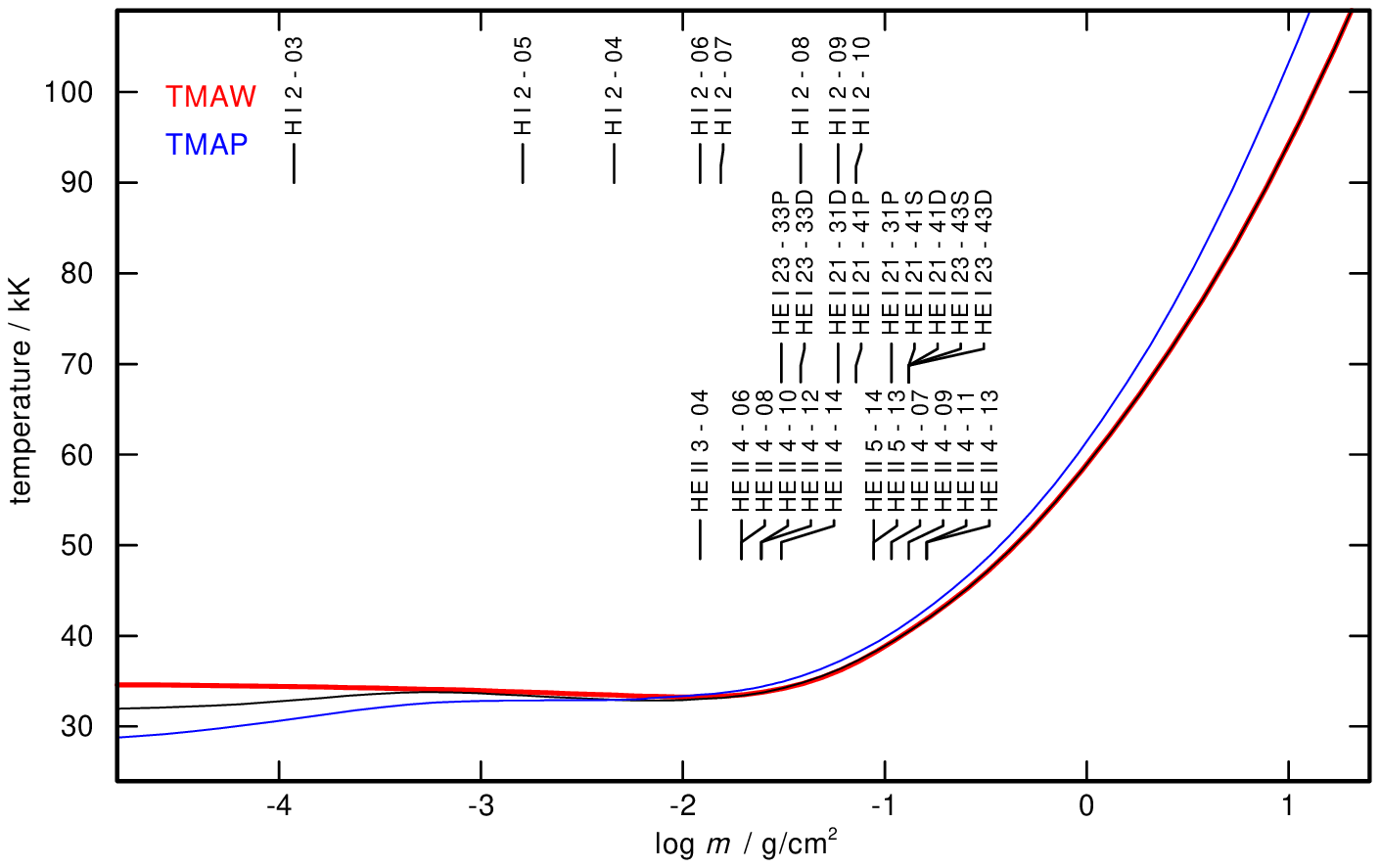}
\caption{Comparison of the \emph{TMAP} (thin, blue) and \emph{TMAW} (thick, red) temperature structure
         (\Teffw{42000}, \loggw{5.46}). 
         The black line shows the structure after the post-calculation steps. 
         The formation depths of the line cores of hydrogen and helium are marked.}
\label{fig:t-structure}
\end{figure}

To ensure a reasonable calculation time, \emph{TMAW} uses adjusted, relatively small model atoms 
(e.g\@. 207 NLTE levels and 498 lines, Sect\@.\,\ref{sect:tmaw}) and the Uns\"old-Lucy 
temperature correction \citep{unsoeld_1968}. This temperature correction is numerically stable, 
but the structure differs from one calculated with standard corrections (Fig.\,\ref{fig:t-structure}). 
Although both structures mainly agree in the line-forming regions, the deviations are not
negligible. At $\log m\,\sga\,0$, the temperature structures differ more strongly (Fig.\,\ref{fig:t-structure})
due to the iron-group opacities that are considered in the \emph{TMAP} models. Iron-group opacities
will be considered by \emph{TMAW} in the near future.

\begin{figure}[ht]
\setlength{\textwidth}{13.85cm}
\plotone{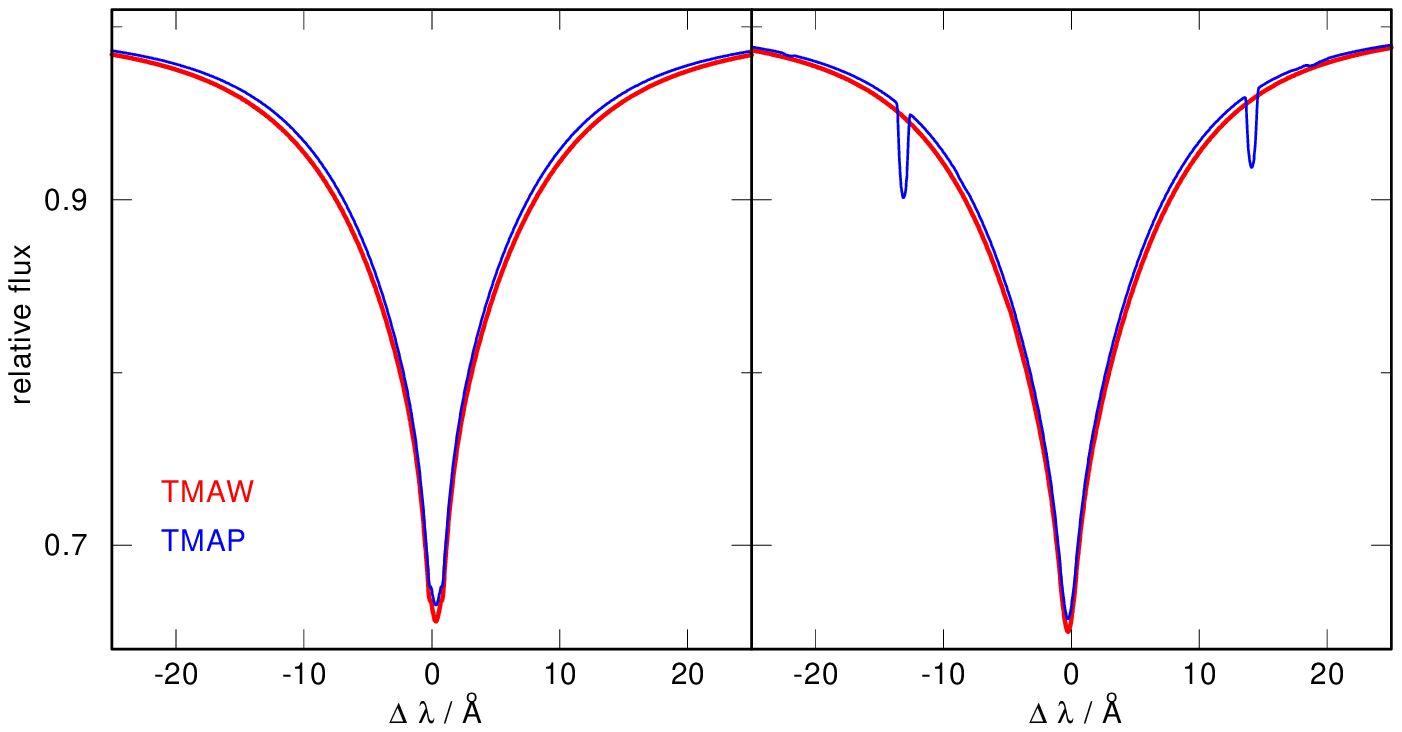}
\caption{Comparison of theoretical H$\beta$ (left) and H$\delta$ (right) line profiles
         calculated from \emph{TMAP} (thin, blue) and \emph{TMAP} (thick, red) models
         (\Teffw{42000}, \loggw{5.46}). 
         H$\beta$ (left) and H$\delta$ (right) are displayed because they are formed at 
         different depths of the atmosphere (Fig.\,\ref{fig:t-structure}). 
         The \emph{TMAP} model shows 
         additional metal lines due additional species included. 
         A rotational velocity of $30\,{\mathrm{km}}/{\mathrm{sec}}$ is applied.}
\label{fig:vgl}
\end{figure}

To compensate this deficiency, we included additional calculation steps, first using the standard
temperature correction applied by \emph{TMAP} and then increasing the number of atomic 
levels in a subsequent line-formation step by using far more detailed model atoms 
(402 NLTE levels and 1652 lines in this example).
The result is shown in Fig.\,\ref{fig:vgl}. With these improvements, the accuracy is improved 
and the calculation time rises only slightly.

\clearpage

\acknowledgements This work is supported by the Federal Ministry of  Education and Research (BMBF) 
under grant 05A11VTB. 
The UVES spectra used in this analysis were obtained as part of an ESO Service Mode run,
proposal 66.D-1800.
The \emph{TMAW} service (http://astro-uni-tuebingen.de/~TMAW) used to calculate
theoretical spectra for this paper was constructed as part of the activities of
the German Astrophysical Virtual Observatory.

\vfill

\bibliography{ringat}

\end{document}